# Current status of data center for cosmic rays based on KCDC[†]

## V.A. Tokareva[1, a], D.G. Kostunin[1], A. Haungs[1]

[1] *Institute for Nuclear Physics, Karlsruhe Institute of Technology, Hermann-von-Helmholtz-Platz 1, Eggenstein-Leopoldshafen, 76344, Germany*

E-mail: [a] victoria.tokareva@kit.edu

We present a current status of data center based on KCDC (KASCADE Cosmic Ray Data Centre), which was originally designed for providing an open access to the events measured and analyzed by KASCADE-Grande, a cosmic-ray experiment located in KIT, Karlsruhe. In the frame of the German-Russian Astroparticle Data Life Cycle Initiative we extend KCDC in order to provide an access to different cosmic-ray experiments and make possible aggregation and joint querying of heterogeneous air-shower data. In the present talk we discuss the description of data and metadata structures, implementation of data querying and merging, and first results on including data of experiments located in Tunka, Russia, in this common data center.

---

[†] financially supported by Russian Science Foundation and Helmholtz Society, grant No. 18-41-06003

# 1. Introduction

Nowadays great results in the field of astroparticle physics are achieved because of a rapid evolution of detector electronics and algorithms and modern computational methods (distributed computing, machine learning), which has come in use in recent years. According to Astroparticle Physics European Coordination committee (APPEC), the common data rate for astrophysical experiments all together is a few PBytes/year, which is comparable to the current LHC output [1]. Thence, from one side we face with a big amounts of data, and from the other side the growing complexity of the computations we would like to perform. To achieve the stable reliable work of data processing at every stage, the proper data life cycle management is required. Thereby we discuss organizing such a cycle in a framework of German-Russian astroparticle data life cycle initiative and discuss issues of the data integration and data workflow management.

**1.1. The project's aim and objectives**

The aim of the initiative is implementing data life cycle, i.e. the data management pipeline, which would suit well for the specific requirements we face in the field of astroparticle physics today, to name the major:

- Relatively big (from 100 to 1000 Tb/year) amounts of data modern astroparticle experiments have to collect in order to observe the messengers of our interest in the most precise way;
- Relatively small amount of events, which are in high interest of us;
- Interest in joint data analysis for statistics increasing;
- Potentially different structure of data, accumulated with different types of detectors, for example, for the high energy cosmic ray data collected with Cherenkov water tank detectors and data about the same messenger taken with scintillator arrays;
- General high demand for open access data sharing for research, outreach and education, which is rapidly gaining ground in particle physics nowadays.

For gaining the aim proposed we develop the common data life cycle for two astroparticle physics projects KASCADE-Grande [2] and TAIGA [3], organized for at studying high-energy cosmic rays by observing extensive air showers (EAS).

**1.2. KASCADE-Grande experiment**

The KASCADE [2] experiment has definitely become a significant milestone in astroparticle physics. It was started in the beginning of 90-s and aimed at studying high energy cosmic rays by observing extended air showers. The setup consisted of several detectors: scintillation counters of various precision, in particular, less precise but more numerous KASCADE detectors and more advanced Grade detectors; an hadronic calorimeter; and a digital radio array LOPES. The experiment has been running for more than 10 years and has collected important data. It's analysis has already led to several notable physical results (see [4−6] and refs. therein) and is still ongoing.

KASCADE-Grande is the first and still the only big cosmic ray experiment that has completely published it's data. Specialized web-portal KCDC [7] has been developed for this purpose and we suppose that it's infrastructure could become a good basis for joint data access and analysis for both the experiments and the possible extensions.

**1.3. TAIGA experiment**

TAIGA [3] detectors are located in Russia in Tunka Valley near Baikal lake. It consists of various detector types, including: Cherenkov photomultipliers of various sensitivity; scintillation counters; a radio detector array; and modern Imaging Air cherenkov Telescopes (IACT). The setup is currently operating and still being extended.

The combined data analysis from all these detectors is already a challenging task, but we decided to start with a bit different approach and perform a combined analysis of data from KASCADE-Grande and Tunka-133. Both the detectors have observed cosmic rays, but in different time, in different locations and different atmospheric conditions. One of them consists of scintillating counters, the other one from cherenkov photo multipliers.

The data amount for this task is fairly limited but looking forward into the perspective of extending our solution for the whole experiments data we are considering the ways to make our solution scalable enough.

**1.4. KASCADE Cosmic Ray Data Center**

The web portal KCDC [7] provides the access to data of the KASCADE-Grande setup for the interested public, i.e. for professional physicists from astroparticle physics community, lecturers, students, and all broad audience. The IT infrastructure of the portal includes highly-demanded technologies, such as interactive data selections, high-availability message exchanging and dynamic task distribution.

We work on extending KCDC by integrating the TAIGA data into the data center workflow, see Fig. 2. For a fast data search the metadata-based approach described further is proposed.

## 2. Data integration

The problem of data integration is of interest because there is no generally accepted solution in the field of astroparticle physics so far. Besides, since astroparticle physics is an observational science, and one cannot influence the phenomena that generate data, the experiments tend get as much as possible from the data they are able to obtain, which results in a data-driven approach for the data analysis. Both conventional algorithmic analysis and deep learning using neural networks can be used, but the problem of scaling of heterogeneous data arises in both cases.

**2.1 Data storage and retrieval**

The data corresponding to different experiments possess many essential distinctions, that makes the problem challenging. The data format itself is different, as each detector measures its own set of observables. Then, the software for the initial steps of data analysis is strongly detector-depended on the detector and could require special software environment of libraries installed. These differences result in the first problem that is creating a unified software interface for working with data. By the moment the whole analysis procedure is designed separately and incompatibly for both KASCADE and TAIGA experiments. Our goal is to develop the unified interface, where the common analysis steps would be designed in a common way, and provide well-enough encapsulation of low-level details to hide them from end user.

The basic functionality of the data management pipeline is to provide data to scientists on demand. A request is a set of conditions and logical operations on them that determine which data the user wants to receive.

Data storage is going to be organized distributedly by the means of a virtual file system deployed on top of existing servers, the most likely candidate for it that is being considered is CernVM-FS [8].

Since the data size is huge and its structure is diverse, a direct search within the data would be extremely slow and resource-consuming, and thus is not going to be implemented. Fortunately, the data have a common metadata format, which includes time, place, atmospheric conditions, etc. A centralized database containing the metadata of all events from both experiments would be used to process data-retrieval requests. The proposed database structure is presented in Fig. 1. In case any kind of requests using the properties not included in this database prove necessary, the appropriate information must be extracted from the data and inserted into the metadata registry.

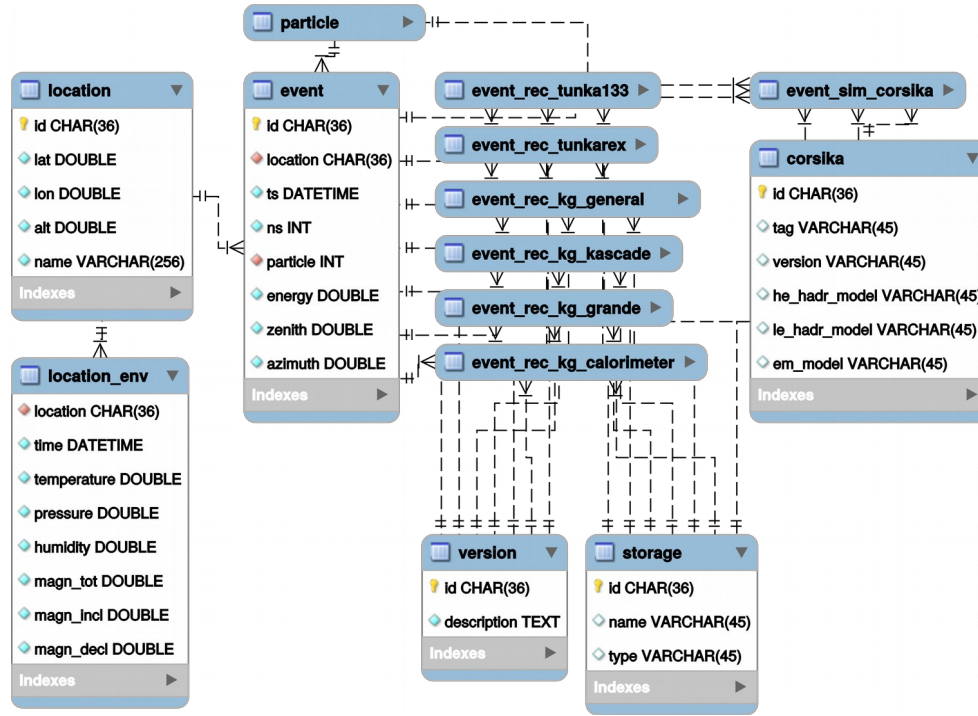

Fig. 1. Proposed design of a database containing event metadata.

### 2.2 Data analysis

Besides the delivery of requested data, the task of data analysis requires access to computer facilities and software for the analysis. Data analysis generally includes two types of sub-tasks: simulations and analyzing experimental data.

Simulation for the physics of air showers consists of simulating two consecutive processes: first, the shower propagation in the atmosphere, and second, the detector response. Both ones are performed independently for each event, require a lot of computing power and produce data volumes comparable with that of experimental data. Thus, parallel computing is a natural solution in this case. Simulations of a shower uses standard CORSIKA software [9] for any experiment and does not depend on detector features. Instead, detector simulations require the software that is usually developed separately for each particular experimental setup, and a special computing environment for it, like libraries or OS versions.

The same requirement of a dedicated environment applies to the software for analyzing experimental data. Distributed execution of such type of tasks could be organized using virtualization techniques, when an image or container with all necessary software is deployed at each processing node.

The demands to keep the uniform CPU load for the computing nodes and to minimize the data transfer overheads result in the necessity to employ a workload management system (WMS) for the distributed analysis. We are going to undertake a comparative performance analysis for various WMS in order to find the one most suitable for the extended KCDC portal integrating KASCADE and TAIGA data.

The workflow scheme for the joint data analysis is presented in Fig. 2 and includes the following steps. First, data acquisition is being done for each experiment, resulting in a set of registered events. For each event dedicated simulations are performed. These data are used to reconstruct the shower properties and the characteristics of an initial particle.

In order to perform a further joint analysis of events from different experiments, data mapping is required, that consists of comparing preliminary distributions of various observables, finding a common scale and normalizing the data according to it, followed by a new iteration of data reconstruction when necessary.

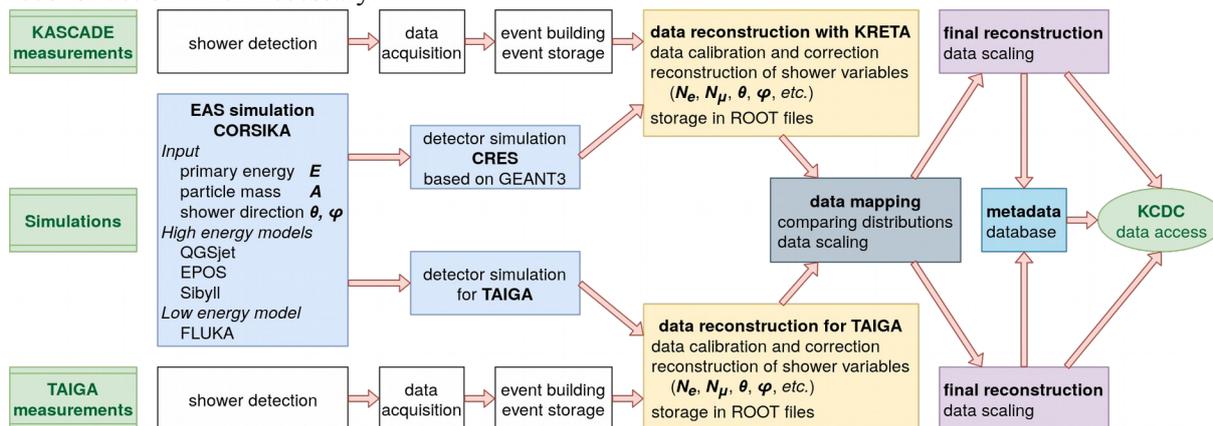

Fig. 2. Proposed scheme of the workflow for the joint analysis of KASCADE and TAIGA data.

## 3. Conclusion

The constantly growing amount of accumulated astroparticle data and the request for the multi-messenger astronomy and machine learning, enable us to develop a unified system for astroparticle data storage and processing.

In a framework of German-Russian astroparticle data life cycle collaboration we proposed the concept of astroparticle data center based on KCDC. Our plans include organization of distributed data storing and processing, creating a platform for joint data analysis for both experiments and providing the data and analysis results to the public open access, as well as using them for the educational and outreach activity. With taking into account the specific features of the research field and keeping in mind the data-oriented approach, we proposed the structure of the metadata database and the possible data workflow scheme.

The built-up infrastructure is supposed to be used to analyze combined data sets with large statistics, allowing to study galactic sources of high-energy γ-rays, which could be a notable step forward in multi-messenger astroparticle physics.